\documentclass[prl,twocolumn,showpacs,preprintnumbers,amsmath,amssymb]{revtex4}
\usepackage{dcolumn}
\usepackage{bm}
\usepackage{graphicx}
\usepackage{bm}
\begin{document}
\pacs{75.47.Lx, 75.40.Cx, 71.38.Ht}
\title{Effect of Al substitution on the charge ordered Pr${_{0.5}}$Ca${_{0.5}}$MnO${_3}$: Structure, Magnetism and Transport}
\author{Sunil Nair and A. Banerjee}
\affiliation{Inter University Consortium for D.A.E. Faclilites,\\ University Campus, Khandwa Road, Indore, 452 017, INDIA.}
\date{\today}
\begin{abstract}
We report the effect  of Mn site substitution of Al on the structural, magnetic and transport properties of charge ordered Pr${_{0.5}}$Ca${_{0.5}}$MnO${_{3}}$.  This substitution introduces impurities in the Mn-O-Mn network, causes the Mn${^{3+}}$/Mn${^{4+}}$ ratio to deviate away from unity and is seen to predominantly replace Mn${^{4+}}$.  The strength of charge ordering is supressed with increasing Al doping, as is seen from ac susceptibility, the magnetocaloric effect and dc resistivity.  The observation of traces of charge ordering over an extended range of impurity doping of 10\% besides being indicative of the robust nature of charge ordering in Pr${_{0.5}}$Ca${_{0.5}}$MnO${_{3}}$, also underscores the relatively insignificant effect that Al substitution has on the lattice distortions and the importance of non-magnetic substitutions.  For all the samples, the conductivity in the temperature range above the charge ordering transition is observed to be due to the adiabatic hopping of small polarons, whereas below the charge ordering transition, Mott's variable range hopping mechanism is seen to be valid. 
\end{abstract}
\maketitle
Mn site doping in half doped charge ordered(CO) manganites has emerged as a popular means of understanding the CO transition, tailoring its strength, and inducing various interesting phases in this class of materials\cite{hardy,maig,rev1}.  Doping in the Mn site besides introducing random impurities in the electrically and magnetically active Mn-O-Mn network, causes the Mn${^{3+}}$/Mn${^{4+}}$ ratio to deviate away from unity hence destabilising the CO. In some cases where the dopants have partially filled d bands, doping is known to broaden the bandwidth, thus causing the ground state to change from an Antiferromagnetic insulator (AFMI) to a Ferromagnetic metallic (FMM) one\cite{hebert}.  The prediction (and experimental verification) of microscopic and mesoscopic Phase seperation(PS)\cite{uehara,rad} has added considerably to the interest in these compounds.  There are calculations\cite{jer} which show that $\leq$ 50\% of hole doping makes these systems unstable towards PS into FMM and CO insulating phases, making the studies of systems around half doping particularly relevant.   

Here we report structural, magnetic and transport properties of Al doping in the Mn site of half doped Pr${_{0.5}}$Ca${_{0.5}}$MnO${_{3}}$, concentrating on the region around the CO temperature.  Al as a dopant is particularly attractive as is clearly seen in Fig. 1 which shows the ionic radii and the number of electrons in the d shell of all the Mn site dopants reported till date.  Its ionic radii matches very well\cite{shan} with that of Mn${^{4+}}$ and hence one would expect a preferential replacement of Mn${^{4+}}$ with increasing Al doping.  This would also cause very little lattice distortion, an aspect which is very important considering the influence of the lattice distortions on the magnetism and transport in this class of materials.  Also, the fact that Al with an empty d band is non magnetic would ensure that unlike popular dopants like Cr\cite{kim} and Fe\cite{levy} which are known to drive the system towards a ferromagnetic ground state, Al doping would result in a more systematic reduction in the strength of the magnetic interactions driving the CO transition. 

Polycrystalline samples of the series Pr${_{0.5}}$Ca${_{0.5}}$Mn${_{1-x}}$Al${_x}$O${_3}$ with $0 \leq x \leq 10\% $ have been prepared by the standard solid state ceramic route with starting materials Pr${_6}$O${_{11}}$, CaCO${_3}$, MnO${_2}$ and Al${_2}$O${_3}$ of alteast 99.99\% purity.  The powdered samples were sintered at 950\raisebox{1ex}{\scriptsize o}C twice for 24 hours each. After regrinding, the samples were pelletised and treated at 1250\raisebox{1ex}{\scriptsize o}C for 36 hours and at 1350\raisebox{1ex}{\scriptsize o}C for 50 hours, with intermediate grindings.  The X-ray Diffraction(XRD) measurements were done using a Rigaku Rotaflex RTC 300RC powder diffractometer with Cu K$\alpha$ radiation. All the samples were seen to be single phase, and the XRD patterns were analysed by the Rietveld profile refinement, using the profile refinement package by Young \textit{et al} \cite{dbws}.  Iodometric Redox titrations using sodium thiosulphate and pottasium iodide were done to estimate the Mn${^{3+}}$/Mn${^{4+}}$ ratio.  Low field AC susceptibility was measured using a home made susceptometer\cite{ashna}. MH isotherms measured using a home made Vibrating sample magnetometer \cite{krsna} were used for determining the magnetocaloric effect.  Dc resistivity measurements were done in the standard 4-probe geometry.

XRD patterns for all the samples are shown in Fig.2. All the diffractograms could be indexed to the orthorhombic $\it{pnma}$ space group.  Table 1 summarises the relevant structural parameters obtained by Rietveld analysis of powder XRD data.  As is seen, the variation in the structural parameters are of the order of 1\%  or less, indicating that Al doping does not introduce any major structural distortion.  The \textit{goodness} of the fits can be gauged by the ratio of  R${_{wp}}$/R${_{exp}}$ which is of the order of 1.15 for all the samples\cite{young}.  The values of the mean Mn valence determined using iodometry indicates that on doping, Al predominantly replaces Mn${^{4+}}$ as was seen earlier in Al substituted LaMnO${_3}$\cite{krsna1}(Table 1).
\begin{table}
\caption{\label{tab:table 1}Structural and fitting parameters determined from the Rietveld profile refinement of the powder XRD patterns for the series Pr${_{0.5}}$Ca${_{0.5}}$Mn${_{1-x}}$Al${_x}$O${_{3+\delta}}$.  The mean valence state of Mn was determined by redox iodometric titrations}
\begin{ruledtabular}
\begin{tabular}{cccccc}
Sample &$x=0\% $ &$x=2.5\% $ &$x=5\% $ &$x=7.5\% $ &$x=10\% $\\
\hline
a($\AA$) &5.4057 &5.3993 &5.3948 &5.3894 &5.3836\\
b($\AA$) &7.6138 &7.6101 &7.6055 &7.5962 &7.5912\\
c($\AA$) &5.3954 &5.3920 &5.3892 &5.3859 &5.3836\\
V($\AA{^3}$) &222.06 &221.55 &221.11 &220.49 &220.01\\
$<$Mn-O$>(\AA)$  &1.942 &1.943 &1.939 &1.936 &1.932\\
$<$Mn-O-Mn$>$  &158.36\raisebox{1ex}{\scriptsize o} &158.43\raisebox{1ex}{\scriptsize o} &158.99\raisebox{1ex}{\scriptsize o} &158.71\raisebox{1ex}{\scriptsize o} &158.54\raisebox{1ex}{\scriptsize o}\\
R${_{wp}}$ &19.35 &18.82 &18.58 &18.82 &18.64\\
R${_{exp}}$ &16.79 &16.37 &16.13 &16.56 &16.34\\
Mn${^{3+}}$\%  &49.4 &50.3 &48.8 &48.4 &46.9\\
Mn${^{4+}}$\% &50.6 &47.2 &46.2 &44.1 &43.1\\
\end{tabular}
\end{ruledtabular}
\end{table}

The magnetic susceptibility is reported to drop in magnitude when cooled across the charge ordered transition in Pr${_{0.5}}$Ca${_{0.5}}$MnO${_3}$\cite{zirak}. This feature can be explained qualitatively to arise due to the structural transition driven by the ordering of the d${_{z{^2}}}$ orbitals of the Jahn Teller distorted Mn${^{3+}}$ ion.  Localisation of the conduction electrons which drive ferromagnetism in these materials result in the ferromagnetic spin fluctuations to be replaced by antiferromagnetic ones, when cooled across the charge ordering transition.  This has been experimentally observed through inelastic neutron scattering measurements on a similar charge and orbital ordered system Bi${_{1-x}}$Ca${_x}$MnO${_3}$\cite{bao}.

Fig. 3 shows the real part of the ac susceptibility ($\chi{_1}$) measured for the series Pr${_{0.5}}$Ca${_{0.5}}$Mn${_{1-x}}$Al${_x}$O${_3}$. The parent compound  Pr${_{0.5}}$Ca${_{0.5}}$MnO${_3}$ is reported to undergo a charge ordering transition at $\approx$245K and a long range antiferromagnetic transition at $\approx$170K\cite{zirak}.  As is seen in Fig. 3, the feature in susceptibility described above is clearly seen in the form of a peak in the first ordered susceptibility($\chi{_1}$).  This feature is seen to decrease in magnitude and shift to lower temperatures with increasing Al doping, clearly indicating the weakening of the CO state.  It is interesting to note that while previous reports have shown the CO state to be destroyed by small ($\sim 5\% $ or less) amounts of impurity doping\cite{kim,levy}, in our case traces of CO can be seen till upto 10 \% of Al doping.  This is a reflection on both, the robustness of the CO in the parent  Pr${_{0.5}}$Ca${_{0.5}}$MnO${_3}$, as well as the fact that Al does not introduce either lattice distortions or magnetic interactions of its own.  Thus a gradual reduction in the CO strength is achieved using Al doping.  \textit{To the best of our knowledge this is the first observation of CO in a system with upto 10\%  of impurity doping at the Mn site}.  

The supression of CO as seen in the low field ac susceptibility measurements were reconfirmed using the measurements of the magnetocaloric effect (MCE) determined by using MH isotherms.  MCE refers to the change in the isothermal magnetic entropy$|\Delta S{_M}|$ \cite{tishin} produced by changes in the applied field, and is known to be appreciable in manganites across the CO transition\cite{sand}.  For measurements made at discrete field and temperature values, it is given by  
\[
|\Delta S{_M}| = \sum\frac{(M{_n}-M{_{n+1}}){_H}}{T{_{n+1}}-T{_n}} \Delta H{_n}
\]
where M${_n}$ and M${_{n+1}}$ are the magnetisation values measured at field H${_n}$ at temperatures T${_n}$ and T${_{n+1}}$ respectively.  Fig. 4 shows the temperature variation of $|\Delta S{_M}|$ for the series Pr${_{0.5}}$Ca${_{0.5}}$Mn${_{1-x}}$Al${_x}$O${_3}$.  The magnitude of the magnetic entropy change associated with the CO transition\cite{sand} is seen to reduce with Al doping, clearly indicating a gradual supression of the CO strength with increasing Al doping.  It is to be noted that the temperatures of the peak in $|\Delta S{_M}|$ matches well with the temperatures where the feature in ac susceptibility is seen, and hence can be denoted as T${_{CO}}$.

It is generally accepted\cite{sal} that the high temperature conductivity in manganites occurs through the hopping of charge carriers localised in the form of small polarons.  These small polarons are known to form in systems with strong electron-phonon interactions, where the charge carrier is susceptible to self localisation through energetically favorable lattice distortions.  In our case, this lattice distortion would be the Jahn-Teller distortion on the Mn${^{3+}}$ lattice site. If the lattice distortions are slow as compared to the charge carrier hopping frequencies, then the hopping is adiabatic and the conductivity can be written as 
\[
\sigma = ne\mu =\frac{3}{2}\frac{ne{^2}a{^2}\nu }{k{_B}}\frac{1}{T} exp (\frac{-W{_H}}{k{_B}T})
\]
where $n$ is the polaron density, $e$ the electron charge, $\mu$ the polaron mobility, $a$ the hopping distance, $k{_B}$ the Boltzmann constant, $\nu$ the effective frequency at which the carrier tries to hop to a neighbouring site and $W{_H}$ the hopping energy.  Thus, conductivity measurements fitted to a ln($\sigma T$) vs 1/T plot should lead to a straight line.  Fig. 5 shows this fitting for the series Pr${_{0.5}}$Ca${_{0.5}}$Mn${_{1-x}}$Al${_x}$O${_3}$.  Good fits could be obtained for all the samples in the region T$>$ $T{_{CO}}$, and the temperatures at which a deviation from the fit is seen matches well with the temperatures at which the feature associated with CO is seen in the magnetic measurements.  It is interesting to note that conductivity in the paramagnetic region of the half doped layered compound LaSr${_2}$Mn${_2}$O${_7}$ has recently been ascribed to be due to the variable range hopping of polarons in the presence of a Coulomb gap\cite{chen}.  However, our results differ, probably due to the difference in the nature of the samples used and is seen to match well with data available on other 3D half doped systems\cite{quen}. Table 2 summarises the values determined from the fitting procedure.  The activation energy for hopping is seen to reduce with Al doping.  The change in the effective polaronic density  is a reflection on the variation of the percentage of Mn${^{3+}}$ as a function of Al doping.  

\begin{table}
\caption{\label{tab:table 2}Parameters determined from fitting the conductivity data on both sides of the CO transition for the series  Pr${_{0.5}}$Ca${_{0.5}}$Mn${_{1-x}}$Al${_x}$O${_3}$.  The conductivity data for T$>$T${_{CO}}$ is fitted to an adiabatic small polaron model, whereas the conductivity in the region T$<$T${_{CO}}$ is fitted to a 3D Variable range hopping mechanism.}
\begin{ruledtabular}
\begin{tabular}{cccccc}
Sample &$x=0\% $ &$x=2.5\% $ &$x=5\% $ &$x=7.5\% $ &$x=10\% $\\
\hline
W${_H}$(meV) &185.76 &182.80 &180.72 &179.20 &176.24\\
n(10${^{21}}$ cm${^{-3}}$) &2.225 &2.270 &2.207 &2.195 &2.132\\
T${_0}$(10${^8}$ K) &7.593 &3.321 &3.13 &2.143 &2.074\\
\end{tabular}
\end{ruledtabular}
\end{table}
The functional form of conduction in the charge ordered region is far more complicated, and remains to be understood. Close to the CO transition, our transport data is seen to fit to Mott's formula\cite{mott} for the variable range hopping (VRH) mechanism in 3 dimensional systems, where the conductivity is given by
\[\sigma = \sigma{_0} exp (T{_0}/T){^{-1/4}}\]
where T${_0}$ is the Mott's activation energy (in units of K) and can be expressed as 
$T{_0}\approx \frac{21}{k{_B}N(E)\xi {^3}}$
where k${_B}$ is the Boltzmann's constant, N(E) is the density of states and $\xi$ is the localisation length.  A semilog plot of $\sigma$ vs T${^{-0.25}}$ is shown in Fig. 6 for all the samples.  The resistance (and the range of fit) is seen to decrease with increasing Al doping, indicating a destabilisation of the insulating CO state.  A decrease in the value of T${_0}$ with doping would imply an increase in the localisation length ($\xi$), provided the density of states N(E) does not change.  Interestingly, charge ordered magnetite is also reported to show a similar VRH like behaviour below the Verwey transition\cite{gupta}.  The conductivity data when fitted to the variable range hopping model does not fit the whole temperature range below T${_{CO}}$, indicating that there are other competing interactions at lower temperatures, which increase with increasing Al doping. Detailed analysis of the conductivity in the charge ordered region is in progress and will be dealt with in a subsequent communication.

In summary, polycrystalline samples of Al doped Pr${_{0.5}}$Ca${_{0.5}}$MnO${_3}$ have been synthesized to study the effect of impurity doping on CO. Al is seen to preferentially replace Mn${^{4+}}$ and destabilise the CO state.  However, traces of CO is seen to remain to upto 10\% of Al doping as seen by ac susceptibility and the magnetocaloric effect. The conductivity in the paramagnetic region is seen to be through the adiabatic hopping of small polarons, which on cooling through the CO transition becomes of the VRH type atleast near the CO transition.  Since the collapse of the CO state at large applied magnetic fields is known to offer a promising avenue for acheiving CMR through the melting of charge ordering\cite{tom}, it would be realistic to expect a large magnetoresistance in this series of samples at much lower applied fields.

We are grateful to Dr. N. P. Lalla for help in X-ray diffraction and Mr. Kranti Kumar for assistance in experimental work.

\begin{figure}
\caption{The number of electrons in the d shell plotted as a function of the ionic radii for all the Mn site dopants reported till date.  The dashed lines correspond to the ionic radii of Mn${^{3+}}$ and Mn${^{4+}}$ ions.}
\caption{X-ray diffractograms for the series Pr${_{0.5}}$Ca${_{0.5}}$Mn${_{1-x}}$Al${_x}$O${_3}$.  All the samples crystallize in the orthorhombic phase and the structural parameters derived from the Rietveld profile refinement of the XRD data are given in Table 1. }
\caption{The real part of first order ac susceptibility plotted as a function of temperature for the series  Pr${_{0.5}}$Ca${_{0.5}}$Mn${_{1-x}}$Al${_x}$O${_3}$.  The arrows clearly show the supression of the feature associated with charge ordering as a function of increasing Al doping.}
\caption{Temperature dependence of the magnetocaloric effect $|\Delta S{_M}|$ for the series Pr${_{0.5}}$Ca${_{0.5}}$Mn${_{1-x}}$Al${_x}$O${_3}$ at a field of 1 kOe. The peaks indicate the entropy change associated with the charge ordering which is seen to decrease both in magnitude and in temperature with increasing Al doping.}
\caption{Semilog plot of $\sigma$T vs 1/T for the series Pr${_{0.5}}$Ca${_{0.5}}$Mn${_{1-x}}$Al${_x}$O${_3}$. The lines shows a linear fit to the conductivity data indicating adiabatic small polaronic hopping in the paramagnetic region.  The arrows mark the temperature where the conductivity data starts to deviate from the polaronic fit.}
\caption{Semilog plot of $\sigma$ vs T${^{-1/4}}$ for the series Pr${_{0.5}}$Ca${_{0.5}}$Mn${_{1-x}}$Al${_x}$O${_3}$.  The lines are linear fits to the experimental data indicating that the variable range hoppping mechanism is valid, atleast in the temperature range close to the charge ordering transition.}
\end{figure}
\end{document}